\documentclass[aps,prb,10pt,onecolumn,superscriptaddress,nofootinbib,nobibnotes]{revtex4-1}

\usepackage{lmodern}
\usepackage[T1]{fontenc}
\usepackage[utf8]{inputenc}
\usepackage{graphicx}
\usepackage{amsmath}
\usepackage[unicode=true,
            colorlinks=true,
            linkcolor=blue,
            citecolor=blue,
            urlcolor=blue]{hyperref} 
\usepackage{siunitx}
\usepackage{multirow}
\usepackage{upgreek}

\usepackage{colortbl}
\usepackage{footnote}
\usepackage{amssymb}	
\usepackage{hyperref}
\usepackage[flushleft]{threeparttable}
\usepackage[x11names]{xcolor}

\newcommand{\domark}{%
  \vbox to 0pt{
    \kern-\dp\strutbox
    \smash{\llap{\color{red!90!black}\#\kern0.5em}}
    \vss
  }%
}

\immediate\write18{git rev-parse --short HEAD > \jobname.gitinfo }
\newcommand{\gitrev}{\InputIfFileExists{\jobname.gitinfo}{}{ref not available}}

\begin{document}

\title{Plasmonic antenna coupling to hyperbolic phonon-polaritons for sensitive and fast mid-infrared photodetection with graphene}

\author{Sebasti\'{a}n Castilla} \affiliation{ICFO - Institut de Ci\`{e}ncies Fot\`{o}niques, The Barcelona Institute of Science and Technology, Castelldefels (Barcelona) 08860, Spain}
\author{Ioannis Vangelidis} \affiliation{Department of Materials Science and Engineering, University of Ioannina, 45110 Ioannina, Greece}
\author{Varun-Varma Pusapati} \affiliation{ICFO - Institut de Ci\`{e}ncies Fot\`{o}niques, The Barcelona Institute of Science and Technology, Castelldefels (Barcelona) 08860, Spain}
\author{Jordan Goldstein} \affiliation{Department of Electrical Engineering and Computer Sciences, Massachusetts Institute of Technology, Cambridge, MA 02139, USA}
\author{Marta Autore} \affiliation{CIC nanoGUNE, E-20018 Donostia-San Sebasti\'{a}n, Spain}
\author{Tetiana Slipchenko} \affiliation{Instituto de Ciencia de Materiales de Arag\'{o}n and Departamento de F\'{i}sica de la Materia Condensada, CSIC-Universidad de Zaragoza, 50009 Zaragoza, Spain}
\author{Khannan Rajendran} \affiliation{ICFO - Institut de Ci\`{e}ncies Fot\`{o}niques, The Barcelona Institute of Science and Technology, Castelldefels (Barcelona) 08860, Spain}
\author{Seyoon Kim} \affiliation{ICFO - Institut de Ci\`{e}ncies Fot\`{o}niques, The Barcelona Institute of Science and Technology, Castelldefels (Barcelona) 08860, Spain}
\author{Kenji Watanabe} \affiliation{Advanced Materials Laboratory, National Institute for Material Science, 305-0044, Tsukuba, Japan}
\author{Takashi Taniguchi} \affiliation{Advanced Materials Laboratory, National Institute for Material Science, 305-0044, Tsukuba, Japan}
\author{Luis Mart\'{i}n-Moreno} \affiliation{Instituto de Ciencia de Materiales de Arag\'{o}n and Departamento de F\'{i}sica de la Materia Condensada, CSIC-Universidad de Zaragoza, 50009 Zaragoza, Spain}
\author{Dirk Englund} \affiliation{Department of Electrical Engineering and Computer Sciences, Massachusetts Institute of Technology, Cambridge, MA 02139, USA}
\author{Klaas-Jan Tielrooij} \affiliation{Catalan Institute of Nanoscience and Nanotechnology (ICN2), Barcelona Institute of Science and Technology, Campus UAB, Bellaterra, Barcelona, 08193, Spain} 
\author{Rainer Hillenbrand} \affiliation{CIC nanoGUNE, E-20018 Donostia-San Sebasti\'{a}n, Spain} \affiliation{IKERBASQUE, Basque Foundation for Science, 48013 Bilbao, Spain}
\author{Elefterios Lidorikis} \email{elidorik@uoi.gr} \affiliation{Department of Materials Science and Engineering, University of Ioannina, 45110 Ioannina, Greece} \affiliation{University Research Center of Ioannina (URCI), Institute of Materials Science and Computing, 45110 Ioannina, Greece}
\author{Frank H.L. Koppens} \email{frank.koppens@icfo.eu} \affiliation{ICFO - Institut de Ci\`{e}ncies Fot\`{o}niques, The Barcelona Institute of Science and Technology, Castelldefels (Barcelona) 08860, Spain} \affiliation{ICREA - Instituci\'o Catalana de Recerca i Estudis Avan\c{c}ats, 08010 Barcelona, Spain}

 \begin{abstract}
 \textbf{Integrating and manipulating the nano-optoelectronic properties of Van der Waals heterostructures can enable unprecedented platforms for photodetection and sensing. The main challenge of infrared photodetectors is to funnel the light into a small nano-scale active area and efficiently convert it into an electrical signal. Here, we overcome all of those challenges in one device, by efficient coupling of a plasmonic antenna to hyperbolic phonon-polaritons in hexagonal-BN to highly concentrate mid-infrared light into a graphene $pn$-junction. We balance the interplay of the absorption, electrical and thermal conductivity of graphene via the device geometry. This novel approach yields remarkable device performance featuring room temperature high sensitivity (NEP of 82 pW$/\sqrt{\rm\mathbf{Hz}}$) and fast rise time of 17 nanoseconds (setup-limited), among others, hence achieving a combination currently not present in the state-of-the-art graphene and commercial mid-infrared detectors. We also develop a multiphysics model that shows excellent quantitative agreement with our experimental results and reveals the different contributions to our photoresponse, thus paving the way for further improvement of these types of photodetectors even beyond mid-infrared range.}\\
 \end{abstract}

\maketitle

Hyperbolic phonon-polaritons (HPPs) are hybridized modes of ionic oscillations and light present in polar dielectric materials, such as hexagonal-BN (hBN)~\cite{Caldwell2014, Caldwell2015a, Basov2016a, Low2016a, Giles2018, Hu2019, Foteinopoulou2019a} that show interesting optical properties such as extreme subwavelength ray-like propagation and sub-diffraction light confinement ($\sim\lambda_0/$100)~\cite{Caldwell2014, Nikitin2016a, Tamagnone2018}, among others. In fact, novel nano-optoelectronic platforms can be attained by merging HPPs functionalities with other 2D materials-based devices, such as graphene photodetectors governed by the photothermoelectric (PTE) effect. This mechanism generates a photoresponse in graphene $pn$-junctions\cite{lemme2011, Gabor2011d, song11, Koppens2014, Peng2018, Schuler2018, Castilla2019, Muench2019} driven by a temperature gradient and Fermi level asymmetry across the channel. Nevertheless, one of the limitations of these detectors is the low light absorption of graphene, especially for mid-IR frequencies where the photon energy becomes comparable to the typical doping level of graphene reaching the Pauli blocking regime\cite{basov08, low14}. This is further exacerbated by the small photoactive area of graphene $pn$-junctions~\cite{Tielrooij2018}, limited by the cooling length of the hot carriers (0.5-1 $\mu$m)~\cite{song11, Gabor2011d, Tielrooij2015g, Tielrooij2018, macdonald09}. These limitations can be overcome by exciting HPPs and focusing them towards the photoactive area and consequently absorbing them in graphene. However, efficient exploitation of HPPs for mid-IR photodetection still remains unexplored.~\cite{Woessner2017a, Pons-Valencia2019}\\

In this work, we embed hBN and graphene within metallic antennas in order to couple their plasmonic interactions with HPPs and achieve highly concentrated mid-IR light on a graphene $pn$-junction for sensitive and fast mid-IR photodetection. Our design (depicted in Fig. 1a-c) combines several mechanisms to achieve high field concentration for both incident light polarizations. Specifically, when light is polarized parallel to the bow-tie antenna axis (Transverse Magnetic, TM-polarization, Fig. 1d), it excites its localized surface plasmon resonance (LSPR) spectrally located at $\lambda \approx$ 5-7 $\mu$m (see Supplementary Information). As a result, the antenna concentrates the incoming mid-IR light into its gap that is situated just above the graphene $pn$-junction (i.e. the detector photoactive area\cite{Castilla2019}). At the same time, the near-fields produced within the antenna hot-spot contain high momenta and thus efficiently launch HPPs ascribed to the spectral overlap of the antenna's LSPR with the hBN upper reststrahlen band (RB) range ($\lambda \approx$ 6-7 $\mu$m). These HPPs propagate as guided modes and interfere within the graphene $pn$-junction, producing high absorption across this small localized region. Likewise, when light is polarized perpendicularly to the bow-tie antenna axis (Transverse Electric, TE-polarization, Fig. 1e), it produces strong light concentration in the gap of the H-shaped antenna, acting as the split-gate, ascribed again to its LSPR spectrally located at $\lambda \approx$ 5.5-7.5 $\mu$m (see Supplementary Information). This phenomenon will also launch hBN HPPs at the gate edges, which will be guided and interfered within the photoactive area.\\

\begin{figure*} [t]
	\includegraphics[width=\textwidth]
	{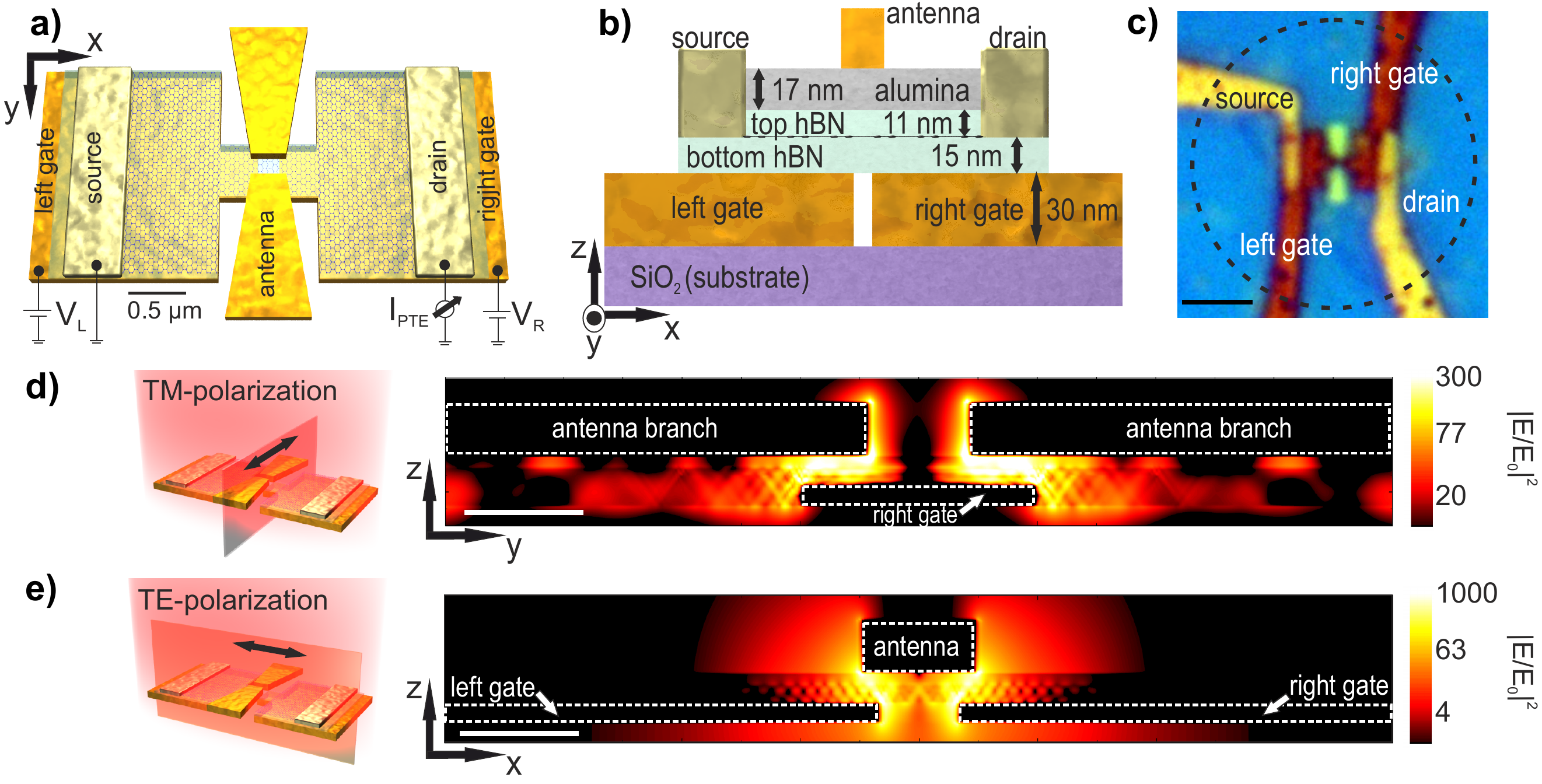}
	\caption{ 
		\footnotesize \textbf{Device schematic and operation principle}. \textbf{a)} Schematic representation of the photodetector consisting of H-shaped resonant gates of 4.2 $\mu$m of total length, with a hBN encapsulated H-shaped graphene channel transferred on top, contacted by source and drain electrodes. A bow-tie antenna of 2.7 $\mu$m of total length is placed on top of the 2D stack. The local gates serve to create a $pn$-junction in the central part of the graphene channel (by applying voltages V${\rm _L}$ and V${\rm _R}$), where the antenna gap and gate gap are located. Both narrow gaps are on the order of $\sim$100 nm. The scale bar corresponds to 0.5 $\mu$m.
		\textbf{b)} Side view of the device design (not to scale) with indications of the materials' thicknesses.
		\textbf{c)} Optical image of the photodetector. The dashed lined circle indicates the typical beam spot size obtained at $\lambda =$ 6.6 $\mu$m. The scale bar corresponds to 2.5 $\mu$m.
		\textbf{d)} Cross section view of the simulated total electric field intensity ($\vert$E$\vert^2$) normalized to the incident one ($\vert$E$_0\vert^2$) along the antenna main axis when light is polarized parallel to the antenna (TM-polarization) axis as indicated in the illustration on the left. The white scale bar corresponds to 250 nm.
		\textbf{e)} Same as \textbf{d} but for light polarization perpendicular to the antenna (TE-polarization) and parallel to the local gates as shown in the schematic on the left.
	}
\end{figure*}



\begin{figure*} [t]
	\centering
	\includegraphics[width=\textwidth]
	{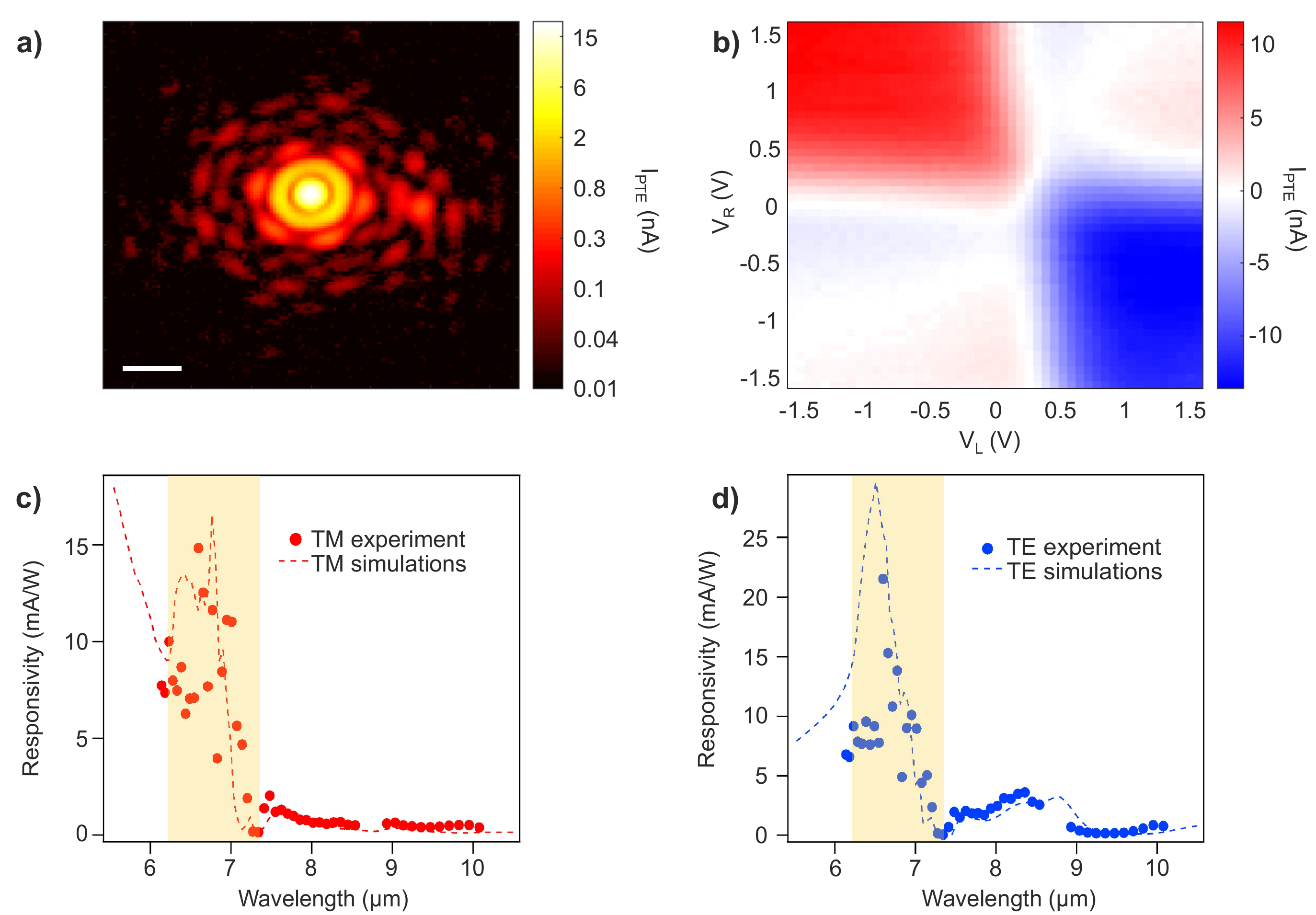}
	\caption{
		\footnotesize \textbf{Photocurrent generation and spectral photoresponse} \textbf{a)} Scanning photocurrent map (log scale) across the mid-IR beam focus at $\lambda$ = 6.6 $\mu$m. The white scale bar corresponds to 20 $\mu$m. We obtain a FWHM of 6.1 $\mu$m. We use a small input power ($\rm P_{\rm in}$) of 13.7 $\mu$W (irradiance of 0.2 $\mu$W/$\mu$m$^2$).
		\textbf{b)} Photocurrent map as a function of the two gate voltages at $\lambda$ = 6.6 $\mu$m. 
		\textbf{c)} Experimental (dots) and theoretical (dashed lines) spectral external responsivity of the device for TM-polarization and \textbf{d)} for TE-polarization. The highlighted region corresponds to the hBN RB ($\lambda$ = 6.2 - 7.3 $\mu$m). For \textbf{c} and \textbf{d}, we set the gate voltages to a $pn$-junction configuration close to the optimal with V$_{\rm L}$ = 0.5 V (97 meV) and V$_{\rm R}$ = -0.5 V (-100 meV). We use the same doping level for the theoretical simulations.
	}
\end{figure*}


To reveal the spatial intensity profile of the beam focus at $\lambda$ = 6.6 $\mu$m, we scan the sample with $xyz$-motorized stages and measure the photocurrent (I$_{\rm PTE}$) as shown in Fig.\ 2a. As a result, we observe the Airy pattern of the beam, which implies that we obtain a well-focused beam (see Methods) and high sensitivity at this wavelength considering the small irradiance input of 0.2 $\mu$W/$\mu$m$^2$. Next, we investigate the photoresponse as a function of the two gate voltages (V$\rm _L$ and V$\rm _R$), shown in Fig.\ 2b, which reveals the photocurrent mechanism and optimal doping level. We find that when sweeping the gate voltages independently, the photocurrent follows several sign changes resulting in a 6-fold pattern, which indicates that the photodetection is driven by the PTE effect, as also shown in other studies in the mid-IR range \cite{Herring2014a, Hsu2015, Woessner2017a}. The highest values of photocurrent occur at $pn$ or $np$ configuration, specifically at V$\rm _L$ = 1.6 V (170 meV) and V$\rm _R$ = -0.82 V (-130 meV), which are relatively low doping levels. We note that when applying a voltage bias in the graphene channel, the photocurrent remains constant while the source-drain current increases linearly with bias (see Fig. S5). This allows us to discard other mechanisms such as photogating and bolometric effects that would increase significantly with voltage bias.\\


\begin{figure*}  [t]
	\centering
	\includegraphics [scale=0.6]
	{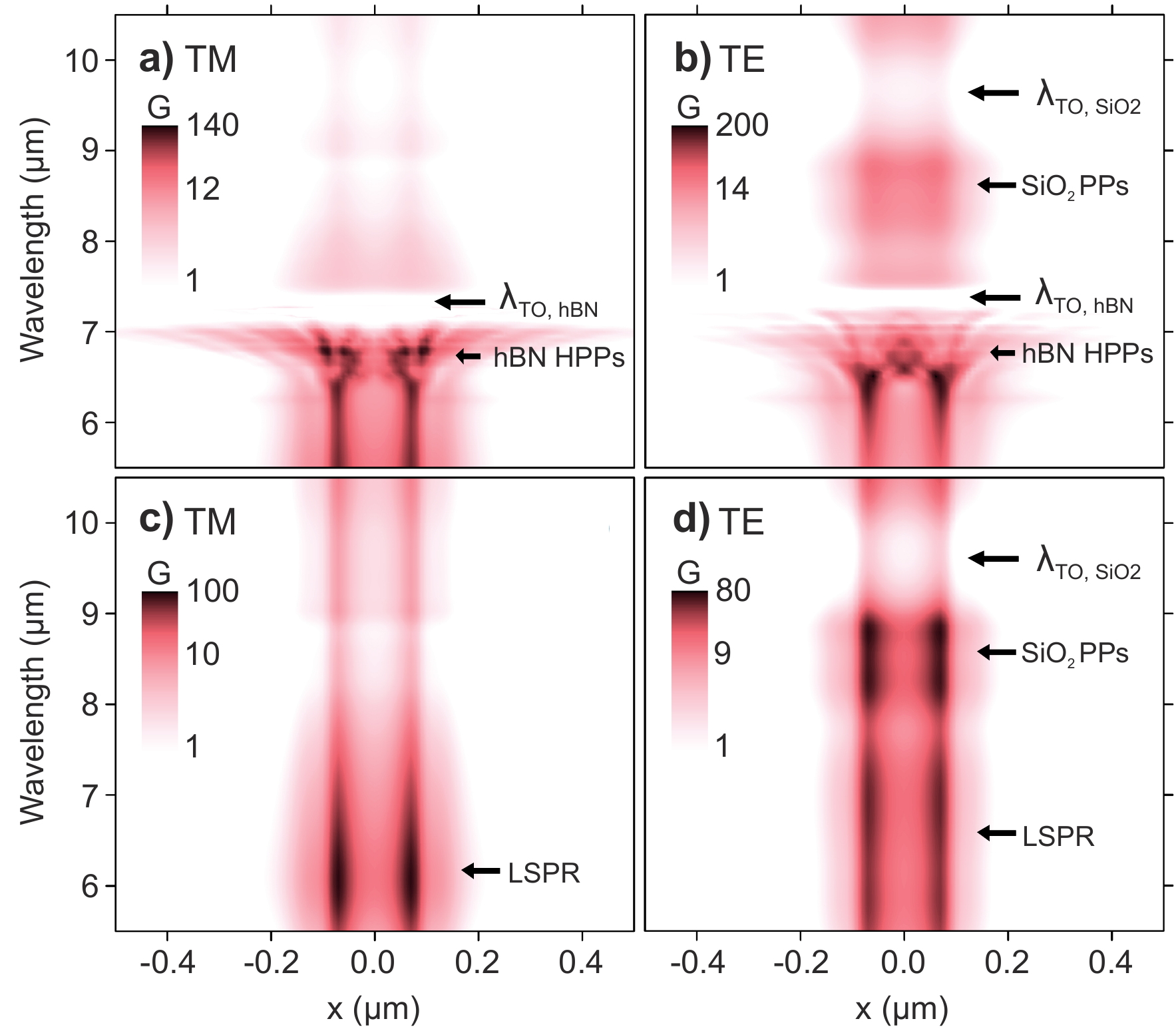}
	\caption{\footnotesize \textbf{Absorption enhancement spectra.} Simulations of the absorption enhancement in graphene (G) along the source-drain direction ($x$ direction as shown in Fig. 1a-b, where x$=$0 is located at the center of the gate gap) as a function of the wavelength, for TM (\textbf{a}) and TE-polarization (\textbf{b}).
		\textbf{c)} and \textbf{d)} correspond to \textbf{a} and \textbf{b} respectively but with wavelength-independent refractive index for hBN ($n$ = 2.4).}\label{label-a}
\end{figure*}



\begin{figure*}  [t]
	\centering
	\includegraphics [scale=0.5]
	{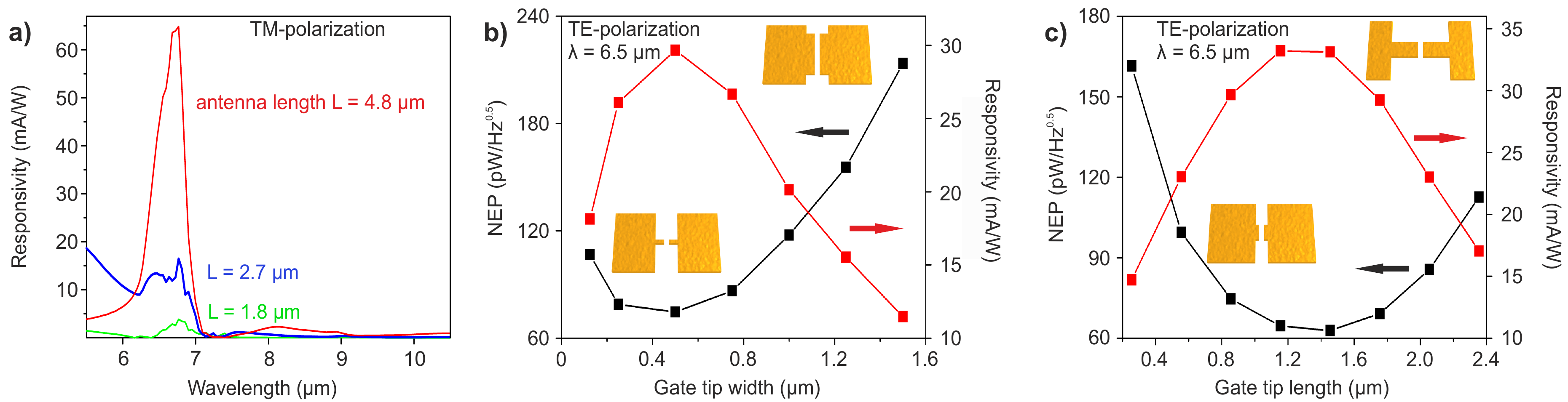}
	\caption{\footnotesize  \textbf{Dependence of the simulated responsivity and NEP on the geometry of the antenna and H-shaped gates}. \textbf{a)} Simulations of responsivity for TM-polarization for different antenna lengths. Different cases are presented: non-resonant antenna within the hBN RB spectral range (antenna total length of L = 1.8 $\mu$m, shown in green), semi-resonant antenna (L = 2.7 $\mu$m shown in blue, which corresponds to the experimental antenna) and resonant antenna (L = 4.8 $\mu$m, shown in red).
		\textbf{b)}  Simulations of responsivity and NEP as a function of gate tip width and graphene (following the exact shape of the gates) as shown in the schematic for TE-polarization at $\lambda$= 6.5 $\mu$m. The tip length is 855 nm, which includes the gap between the gates of 155 nm. The source-drain distance is 2.6 $\mu$m and electrodes width is 2 $\mu$m as in the measured device. 
		\textbf{c)} Same as \textbf{b} but as a function of the gate tip length as shown in schematic. The tip width is 500 nm.
	}\label{label-a}
\end{figure*}


To determine the photodetector spectral response, we measure the TM-polarization (Fig. 2c) external responsivity (see Methods) as a function of excitation wavelength. We obtain high values up to 15 mA/W within 6-7 $\mu$m at the hBN RB. On the other hand, for TE-polarization (Fig. 2d) we observe two responsivity peaks, the first one (up to 22 mA/W) again within the hBN RB (6-7 $\mu$m) and a second peak (3.5 mA/W) around 8 $\mu$m. We also plot the simulated responsivity that is extracted from the multiphysics simulations, which considers the whole device photoresponse (optical excitation, carrier distribution and relaxation, heat diffusion and thermoelectric current collection. See further details in Supplementary Information). We observe excellent qualitative and quantitative agreement between experimental and theoretical responsivity, which we explore in the following by analyzing each component involved in the photoresponse.\\

We first identify the behavior of the resonant mechanisms, in terms of field intensity enhancement and spatial localization by studying the absorption enhancement in graphene (G) across the channel in the $x$ direction (averaging over 500 nm in $y$ direction, see Fig. 1a-b for axis definition) and as a function of the wavelength as shown in Fig. 3. We define G as following: G($\lambda, \textbf{r}$) = $Abs_{device}(\lambda, \textbf{r}) /Abs_{air}(\lambda, \textbf{r})$, which is the ratio between the graphene absorption incorporating all the elements of the device (e.g. antenna, contacts, etc.) to that of suspended graphene as a function of $\lambda$ and the position vector $\textbf{r}$. G and responsivity are proportionally related via the electronic temperature gradient as shown in Fig. S4 and in Supplementary Information.\\



\begin{figure*} [t] 
	\centering
	\includegraphics [scale=0.7]
	{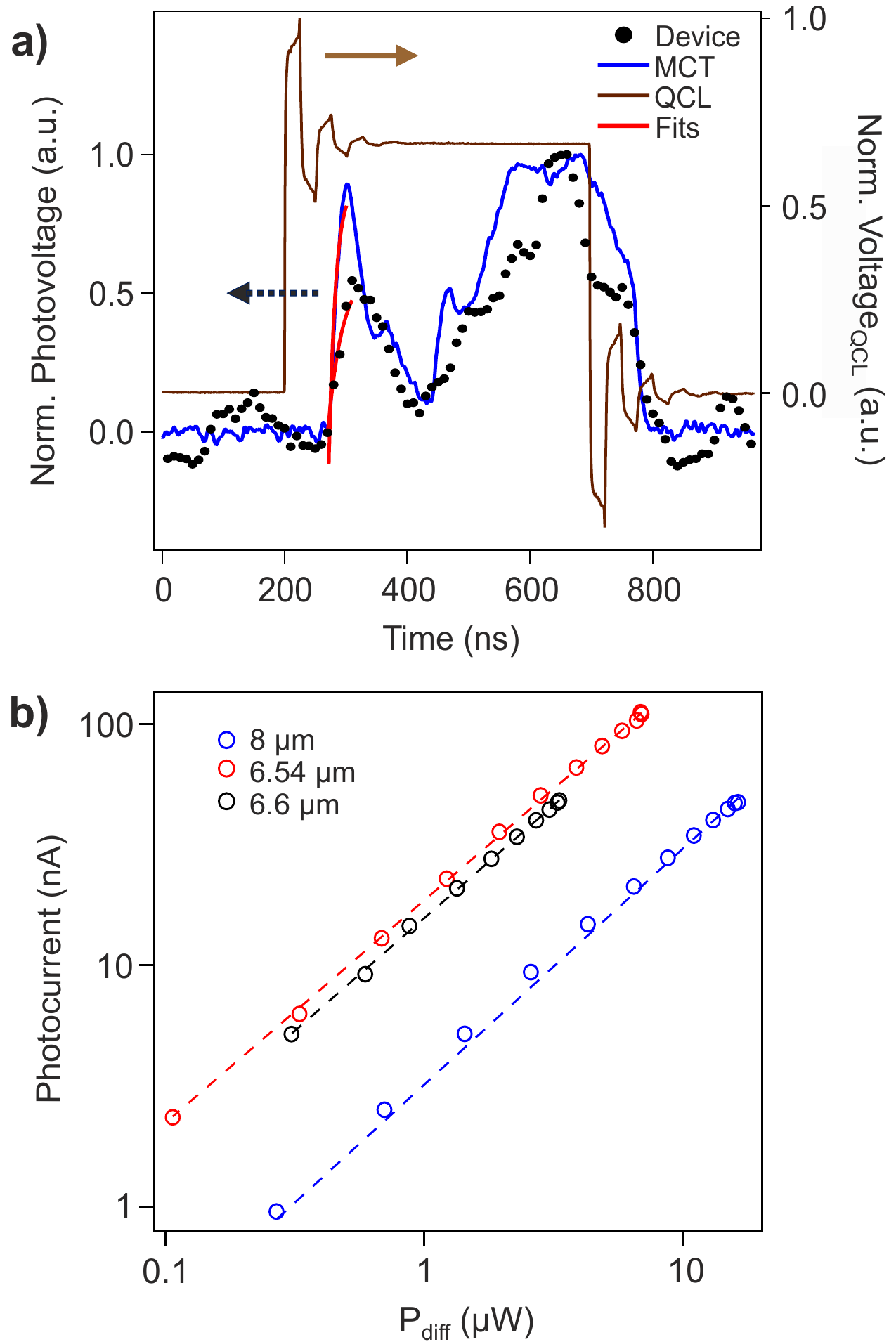}
	\caption{\footnotesize \textbf{Photodetection speed and power dependence}. \textbf{a)} Time-resolved photodetection traces at $\lambda =$ 6.6 $\mu$m, compared with a MCT detector (both plotted in black dots and blue line respectively) and the respective QCL voltage signal (brown line). The QCL pulse width corresponds to 496 ns. The photovoltage fits are shown in red lines. We obtain rise times of 9 $\pm$ 3 ns and 17 $\pm$ 3 ns for the MCT and our device respectively.
		\textbf{b)} Photocurrent as a function of laser power (P$\rm _{diff}$ = P$\rm _{in}\times$A$_{\rm diff}$/A$_{\rm focus}$, see Methods) for different wavelengths on a log-log scale. Circles correspond to the data points, while the dashed lines represent the fits according to I$\rm _{PTE}$ $\propto$ P$^\gamma \rm _{diff}$. Among all cases, $\gamma$ ranges from 0.92-0.98. We observe linear photoresponse over three orders of magnitude of power (limited by the power meter sensitivity range for P$_{\rm in}$ calibration). These results suggest that we are operating in the weak heating regime ($T_e-T_{l}$ $<<$ $T_{l}$) \cite{Tielrooij2015g, Castilla2019}, as in the strong heating regime ($T_e-T_{l}$ $>>$ $T_{l}$), a sublinear behavior is expected ($\gamma =$ 0.5)~\cite{Tielrooij2015g, Castilla2019}. Here $T_{e}$ is the electronic temperature and $T_{l}$ is the graphene lattice temperature, which the latter is in thermal equilibrium with the environment.
	}
	\label{figure_5}
\end{figure*}

In the TM-polarization case shown in Fig. 3a, we observe very high G values at the antenna LSPR ($\lambda \sim$ 6 $\mu$m). The value of G peaks around 6.8 $\mu$m due to the hybridization of the hBN HPPs with the antenna LSPR and to the constructive interference of the propagating HPPs occurring at $x\sim$ +/-100 nm. In fact, the different spatial patterns of G arise from the wavelength dependence of the HPP propagation angle in hBN following the equation $\tan\theta(\omega) = i\sqrt{\varepsilon_{x,y}(\omega)}/\sqrt{\varepsilon_z(\omega)}$ \cite{Caldwell2014, Dai2015b, Woessner2017a}. For longer wavelengths, we find a negligible G between 7-7.3 $\mu$m that corresponds to the hBN transverse optical (TO) phonon. We observe that the highest G values are only found for the spatially confined region (from $x \sim$ -100 to 100 nm) where the antenna and gates overlap, which is designed to coincide with the graphene $pn$-junction. This overlap causes high concentration of the antenna plasmons, HPPs and more efficient reflections of HPPs (see Fig. S6-S7). Nevertheless, in the hBN RB we find large G values outside this tightly localized region due to HPP propagation.\\

For TE-polarization (see Fig. 3b), we find the maximum values of G between 6.2 to 6.6 $\mu$m due to the gate LSPR hybridization with HPPs and their strong constructive interference at $x =$ 0. For longer wavelengths, we identify a G peak centered at 8.5 $\mu$m that corresponds to SiO$_2$ phonon-polaritons (PPs) hybridization with the gate LSPR as presented in Fig. S8-9.\\

To further elucidate the role of the antennas in G, we simulate the system without the contribution of the HPPs using wavelength-independent refractive index values for the hBN (Fig. 3c-d). For TM-polarization (Fig. 3c), we observe a peak around 6 $\mu$m that corresponds to the antenna LSPR and its resonance tail extending up to 8 $\mu$m. For TE-polarization, in contrast, Fig. 3d shows high values of G across a broader wavelength range (5.5-7.5 $\mu$m) due to the complex shape of the gates and their interactions with the source-drain contacts (see Fig. S10). Although in Fig. 3d we observe lower G values compared to Fig. 3c (see also Fig. S8), we obtain higher values of G in TE-polarization when combining the gate LSPR with HPPs (Fig. 3b) ascribed to its higher spectral overlap with the hBN RB and due to the stronger constructive interferences of the HPPs excited by the gates.\\

To evaluate the coupling between the bow-tie antenna LSPR and the hBN HPPs, we study the responsivity as a function of the antenna length for TM-polarization as shown in Fig. 4a (see also Fig. S11). We observe some hBN HPP excitation when using an antenna non-resonant (green line) within the hBN RB range, in which case we obtain a maximum responsitivity of 4 mA/W. In the case of the semi-resonant antenna (experimental antenna, shown in blue line), whose LSPR partially overlaps with the RB spectral range~\cite{Pons-Valencia2019}, the responsivity increases to 17 mA/W respectively. However, this can be significantly improved if we use a longer antenna (red line) such that its LSPR peak fully overlaps with the hBN HPPs peak, thus obtaining 65 mA/W.\\

Next, we examine the impact of the H-shaped gates excited at $\lambda$= 6.5 $\mu$m with TE-polarization on the responsivity and NEP (noise-equivalent power, see Methods) by varying the width and length of the gate tip and graphene as indicated in Fig. 4b-c (see also Fig. S12). Fig. 4b shows that the responsivity (NEP) increases (decreases) when decreasing the tip width down to an optimal value of 500 nm, which coincides with the experimental value. For the case of the gate tip length, the optimum is found around 1.45 $\mu$m, which is relatively close to experimental one (855 nm). These results are ascribed to the balancing act of absorption, electrical resistance and thermal conductivity: larger absorption and lower thermal conductivity increase the temperature gradients, but a smaller electrical conductivity also reduces the photocurrent and thus the responsivity.\\

Now we discuss the technological relevance of our photodetector. First, we measure the photodetection speed by using as reference a commercial fast mercury-cadmium-telluride (MCT) detector. We plot in Fig. 5a the quantum cascade laser (QCL) voltage (brown line) together with the photoresponses of the MCT (blue line) and our device (black circles). The signal of the MCT detector reveals the pulse shape of the laser. We fit an exponential function to the initial peak to determine the rise time (shown in red lines), obtaining a value of 9.5 ns, which is close to its datasheet value of 4.4 ns. In the case of our photodetector, we find a rise time of 17 ns (22 MHz) when using a current amplifier with 14 MHz bandwidth. This suggests that our time-resolved measurements are limited by the current amplifier bandwidth (see Fig. S13), meaning that the actual rise time may be shorter. In fact, our theoretical calculations predict a speed of 53 ps (see Supplementary Information).\\

The sensitivity of the detector is best expressed in terms of external responsivity, which the maximum measured value is 27 mA/W (92 V/W, see Fig. S14), yielding a noise-equivalent-power of 82 pW/$\sqrt{\rm Hz}$~\cite{Woessner2017a, Guo2018a, Cakmakyapan2018, Sassi2017, Yu2018}, assuming the graphene thermal noise as the dominating noise source\cite{Castilla2019, tredicucci12, tredicucci14, chalmers14, chalmers17}. We emphasize that the zero-bias operation leads to low noise levels and a very low power consumption, which is given by the voltage applied to the gates. Furthermore, our novel design allows sensitive detection in different polarizations, which is a limitation for the mentioned graphene detectors~\cite{Woessner2017a, Guo2018a, Cakmakyapan2018, Sassi2017}. Additionally, our device exhibits a wide dynamic range by showing linear photoresponse over three orders of magnitude as shown in Fig. 5b, which is an issue for other types of graphene detectors~\cite{Cakmakyapan2018} and commercial detectors such as MCT~\cite{Rogalski2019}. It also has a very small active area given by the antennas' cross-sections, which implies high spatial resolution and opens the possibility of arranging it into high density photodetector pixels~\cite{Rogalski16, Guo2018a} that are CMOS compatible~\cite{Goossens2017}. All of these performance parameters combined make our device an interesting platform that fulfills the ongoing trend of decreasing the size, weight and power consumption (SWaP) of infrared imaging systems~\cite{Rogalski2019}.\\

The novel device concept introduced in this work can be extended to detectors for other wavelengths or more specific functionalities such as hyperspectral imaging and spectroscopy. Our approach can also be combined with HPPs in other regions of the mid-IR and long-wave infrared range such as MoO$_3$ \cite{Ma2018a, Zheng2018, Zheng2019}. Additional tuning and wavelength sensitivity can be realized by controlling the hyperbolic material’s thickness\cite{DaiS2014, Woessner2017a} or shape\cite{Kalfagiannis2019, Alfaro-Mozaz2019, Caldwell2014, Li2015e}.\\


\section*{Methods}

\subsubsection{Measurements}
We use a pulsed QCL mid-IR laser (LaserScope from Block Engineering) that is linearly polarized and has a wavelength tuning range from $\lambda$ = 6.1 to 10 $\mu$m. We scan the device position with motorized $xyz$-stage. We modulate the mid-IR laser employing an optical chopper at 422 Hz and we measure the photocurrent using a lock-in amplifier (Stanford Research). We focus the mid-IR light with a reflective objective with a numerical aperture (NA) of 0.5. We measure the mid-IR power using a thermopile detector from Thorlabs placed at the sample position.

\subsubsection{Responsivity and NEP calculation}
The external responsivity is given by: Responsivity = (I$_{\rm PTE}/\rm P_{\rm in}$)$\times$(A$_{\rm focus}$/A$_{\rm diff}$)\cite{tredicucci12, tredicucci14, Castilla2019}, where P$_{\rm in}$ is the power measured by the commercial power meter, A$_{\rm focus}$ is the experimental beam area at the measured wavelength and A$_{\rm diff}$ is the diffraction-limited spot size. We measure the photocurrent I$_{\rm PTE}$ from the output signal of the lock-in amplifier $V_{\rm LIA}$ considering I$_{\rm PTE} = \frac{2 \pi \sqrt{2}}{4G} V_{\rm LIA}$\cite{tredicucci12, tredicucci14}, where $G$ is the gain factor in V/A (given by the lock-in amplifier). We use the ratio A$_{\rm diff}/$A$_{\rm focus}$ for estimating the power reaching our photodetector since $A_{\rm diff}$ is the most reasonable value one can attain when considering the detector together with an optimized focusing system (e.g. using hemispherical lens) and it is widely used in the literature for comparing the performances among photodetectors \cite{Castilla2019, tredicucci12, tredicucci14}. We usually have a ratio of A$_{\rm focus}$/A$_{\rm diff}$ $\approx$ 7. This ratio is given by A$_{\rm diff}/$A$_{\rm focus} = \frac{w^2_{\rm 0,diff}}{w_{\rm 0,x} w_{\rm 0,y}}$. In order to obtain $w_{\rm 0,x}$ and $w_{\rm 0,y}$ we use our experimental observation that the photocurrent is linear in laser power and measure the photocurrent while scanning the device in the $x-$ and $y-$direction. Consequently, the photocurrent is described by Gaussian distributions $\propto e^{-2x^2/w^2_{\rm 0,x}}$ and $\propto e^{-2y^2/w^2_{\rm 0,y}}$, where $w_{\rm 0,x}$ and $w_{\rm 0,y}$ are the respectively obtained spot sizes (related to the standard deviation via $\sigma = w_0/2$ and to the FWHM = $\sqrt{2 \ln(2)}w_0$). We usually achieve $w_{\rm 0,x}$ = 5.05 $\mu$m and $w_{\rm 0,y}$ = 5.40 $\mu$m at $\lambda$ = 6.6 $\mu$m (see Fig. S14b). For the diffraction-limited spot, we consider $w_{\rm 0,diff} = \frac{\lambda}{\pi}$, with $\lambda$ the mid-IR laser wavelength. The diffraction-limited area is hence taken as A$_{\rm diff} = \pi w_{\rm 0, diff}^2 = \lambda^2/\pi$. Additionally, the noise-equivalent power (NEP) that characterizes the sensitivity of the photodetector is defined as NEP $ = I_{\rm noise}/$Responsivity and considering that our unbiased photodetector has a very low noise current that is limited by Johnson noise, we use a noise spectral density $I_{\rm noise} = \sqrt{\frac{4k_BT}{R_D}}$, where $k_B$ corresponds to the Boltzmann constant, $T$ is the operation temperature (300 K) and $R_D$ the device resistance. 

\section*{Acknowledgments}
\small
The authors thank David Alcaraz-Iranzo, Jianbo Yin, Iacopo Torre, Hitesh Agarwal, Bernat Terrés and Ilya Goykmann for fruitful discussions. F.H.L.K. acknowledges financial support from the Spanish Ministry of Economy and Competitiveness, through the “Severo Ochoa” Programme for Centres of Excellence in R$\&$D (SEV-2015-0522), support by Fundacio Cellex Barcelona, Generalitat de Catalunya through the CERCA program,  and  the Agency for Management of University and Research Grants (AGAUR) 2017 SGR 1656.  Furthermore, the research leading to these results has received funding from the European Union Seventh Framework Programme under grant agreement no.785219 and no. 881603 Graphene Flagship for Core2 and Core3. ICN2 is supported by the Severo Ochoa program from Spanish MINECO (Grant No. SEV-2017-0706). K.J.T. acknowledges funding from the European Union’s Horizon 2020 research and innovation programme under Grant Agreement No. 804349. S.C. acknowledges financial support from the Barcelona Institute of Science and Technology (BIST), the Secretaria d’Universitats i Recerca del Departament d’Empresa i Coneixement de la Generalitat de Catalunya and the European Social Fund (L'FSE inverteix en el teu futur) – FEDER. T. S. and L.M.M. acknowledge support by Spain’s MINECO under Grant No. MAT2017-88358-C3-1-R and the Aragon Government through project Q-MAD.

\section*{Author contributions}
\small
F.H.L.K., R.H., M.A. and S.C. conceived the project. S.C. fabricated the device and performed the experiments. V.P. assisted in device fabrication and experiments. M.A. supported the device fabrication. I.V. and E.L. performed the simulations and developed the multiphysics model. S.C., T.S. and L.M.-M. assisted in the modelling. S.C., K.R., J.G. and M.A. performed preliminary optical simulations. S.C., I.V., E.L and F.H.L.K. wrote the manuscript. J.G., S.K. and K.-J.T. assisted with measurements and discussion of the results. K.W. and T.T. synthesized the hBN crystals. D.E., R.H., K.-J.T., E.L. and F.H.L.K. supervised the work and discussed the results. All authors contributed to the scientific discussion and manuscript revisions. S.C. and I.V. contributed equally to the work. 


\bibliography{MIRPaper_bib}

\clearpage

\end{document}